\documentclass[aps,pre, twocolumn, showpacs, superscriptaddress, floatfix]{revtex4}

\usepackage{amssymb, amsmath, amsfonts}
\usepackage{multirow}
\usepackage{graphicx}

\begin{document}

\title{A mixed population of competing TASEPs with a shared reservoir of particles} 

\date{19 December 2011}

\author{Philip Greulich$^\dagger$}

\affiliation{SUPA, School of Physics \& Astronomy, University of Edinburgh, James Clerk Maxwell Building, King's Buildings, Mayfield Road, Edinburgh EH9 3JZ, United Kingdom}

\author{Luca Ciandrini$^\dagger$}
\email{l.ciandrini@abdn.ac.uk}

\thanks{\\$\dagger$ denotes equal contribution,$^{\odot}$ also denotes equal contribution}

\affiliation{SUPA, Institute for Complex Systems and Mathematical Biology, King's College,
University of Aberdeen,  Aberdeen AB24 3UE, United Kingdom}

\author{Rosalind J. Allen$^{\odot}$}
\affiliation{SUPA, School of Physics \& Astronomy, University of Edinburgh, James Clerk Maxwell Building, King's Buildings, Mayfield Road, Edinburgh EH9 3JZ, United Kingdom}

\author{M. Carmen Romano$^{\odot}$}
\affiliation{SUPA, Institute for Complex Systems and Mathematical Biology, King's College,
University of Aberdeen,  Aberdeen AB24 3UE, United Kingdom}
\affiliation{Institute of Medical Sciences, Foresterhill, University of Aberdeen, Aberdeen AB25 2ZD, United Kingdom}


\begin{abstract}
We introduce a mean-field theoretical framework to describe multiple  totally asymmetric simple exclusion processes (TASEPs) with different lattice lengths, entry and exit rates, competing for a finite reservoir of particles. We present relations for the partitioning of particles between the reservoir and the lattices: these relations allow us to show that  competition for particles can have non-trivial effects on the phase behavior of individual lattices. For a system with non-identical lattices, we find that  when a subset of lattices undergoes a phase transition from low to high density, the entire set of lattice currents becomes  independent of total particle number. We generalize our approach to systems with a continuous distribution of lattice parameters, for which we demonstrate that measurements of the current carried by a single lattice type can be used to extract the entire distribution of lattice parameters. Our approach applies to populations of TASEPs with any distribution of lattice parameters, and could easily be extended beyond the mean-field case.
\end{abstract}

\maketitle

\section{Introduction}
\label{sec::intro}

The totally asymmetric simple exclusion process (TASEP) has become a paradigm in non-equilibrium statistical physics, due to its rich phenomenology and wide applicability~\cite{schmittman_statistical_1995,derrida_exactly_1998,schutz_exactly_2001,blythe_nonequilibrium_2007,chou_non-equilibrium_2011,chowdhury_physics_2005}.  The majority of this work has focused on the case of a single TASEP with a fixed entry rate (corresponding to infinite availability of particles). However, many physical and biological situations involve multiple lattices with different parameters, whose dynamics is coupled via competition for a finite pool of particles.  In this paper, we present a simple mean-field theoretical framework to study such scenarios, which can be used for an arbitrary number of lattices and for any distribution of lattice parameters. We show that non-trivial behavior, including extension of the shock phase to a finite region of parameter space, and ``buffering'' of individual currents to changes in the reservoir particle number,  can emerge from the competition between lattices for particles. We further show that in these systems, information on the whole system can be extracted from  measurements of the behaviour of an individual lattice subtype.

The standard TASEP is a stochastic process which describes the collective movement of particles along a one-dimensional lattice composed of $L$ sites. On the lattice, a particle hops stochastically from site $i$ to $i+1$, provided that site $i+1$ is empty. Particles enter the first site (if unoccupied)  and leave the last site with fixed rates. This model shows  three distinct phases controlled by the boundary rates: the low density (LD) phase occurs when the entry rate is limiting, the high density (HD) phase when then exit rate is limiting, and the maximal current (MC) phase when both entry and exit rates are large enough that the current is limited only by the hopping rate along the lattice. Exact steady-state solutions for the density of particles on the lattice and the current are known~\cite{derrida_exact_1993,schutz_phase_1993,mallick_2011}.

First introduced as a model of protein synthesis~\cite{macdonald_kinetics_1968, macdonald_concerning_1969}, the TASEP nowadays provides a generic model for transport processes, with applications including  the production of mRNA~\cite{tripathi_interacting_2008, klumpp_stochasticity_2008} and protein~\cite{shaw_totally_2003,dong_towards_2007} in biological cells, the motion of motor proteins along cytoskeletal filaments~\cite{lipowsky_2001,nishinari_intracellular_2005,greulich_intracellular_2007,pierobon_traffic_2009,parmeggiani_non-equilibrium_2009}, the transport of vesicles in fungal hyphae~\cite{sugden_model_2007}, collective insect motion~\cite{john_trafficlike_2009} and traffic flow problems~\cite{chowdhury_statistical_2000}. Most of these studies have focused on particle motion on a single TASEP track (including tracks with multiple lanes, e.g.~\cite{twolane_TASEP_martin});  in reality, however, one often has multiple tracks, which may have different lengths, hopping and boundary rates, and which compete for a finite number of particles. For example, in traffic dynamics vehicles may distribute themselves across different roads, while in biological cells different mRNA molecules  compete for the cellular protein production machinery~\cite{alberts_essential_1998}. Our aim in this paper is to provide a simple, intuitive and generic framework to study such problems.

This work is not the first to consider the effects of a finite reservoir of particles, or of competition for particles between several TASEPs. Recently, Adams {\em{et al}}~\cite{adams_far--equilibrium_2008} and Cook and Zia~\cite{cook_feedback_2009} studied a single TASEP with a finite reservoir of particles, using Monte Carlo simulations, mean-field theory and both simple and generalised domain-wall theories. Cook {\em{et al}}~\cite{cook_competition_2009} extended this work, using Monte Carlo simulations and domain-wall theory to describe several TASEPs of the same or different lengths, sharing the same particle reservoir. This work revealed interesting effects due to the reservoir-induced coupling between TASEPs. In particular, a regime can exist in which reservoir density is independent of the total particle number. The domain-wall approach is  powerful because it can incorporate fluctuations and thus describe accurately the density profiles on the lattices. However, this approach is  difficult to generalize to many lattices with arbitrary distributions of entry and exit rates. Here, we take a much simpler, mean-field approach.  Our approach does not provide information on fluctuations and cannot predict density profiles. However, it yields a simple way to   calculate phase diagrams, and is easy to generalize to arbitrary populations of lattices. 

In section \ref{sec:theory}, we describe our mean-field approach and present relations for the partitioning of particles between the reservoir and lattices. We apply this methodology to the case of a single TASEP coupled to a finite reservoir in section \ref{sec:single} and to a population of identical lattices in section \ref{sec::homo}. As a result of the competition for particles, a new region of the phase diagram emerges, in which the  LD and HD phases coexist; the size of this region  increases as the particle availability decreases. Next, in section \ref{sec::mixed}, we consider a system with two different types of lattices and we find that an LD-HD phase transition on one lattice type causes the behaviour of the other  lattice type to become independent of the particle number. In section \ref{sec::arbitrary},  we generalize to the case of mixed populations of lattices with arbitrary distributions of boundary rates. Here we find that an LD-HD phase transition on any lattice type ``buffers'' all other lattices to changes in the particle number, and we show that the entire distribution of lattice parameters can be extracted from the current on a single lattice subtype. Finally we present our conclusions in section \ref{sec::conclusion}.

\section{Mean-field theory for multiple competing TASEP\lowercase{s}}\label{sec:theory}

\begin{table}	
\begin{tabular}{llll}
$\rho = \alpha$ &\,\,LD phase &\,\,$\alpha<\beta, \alpha< 1/2$ & \,\,$J =  \gamma \alpha \left(1- \alpha \right)$\\
$\rho =  1-\beta$ &\,\,HD phase &\,\,$\alpha>\beta, \beta< 1/2$ &\,\,$J =  \gamma \beta \left(1- \beta \right)$\\
$\rho  =  1/2$ &\,\,MC phase  &\,\,$\alpha,\beta\geqslant 1/2$ &\,\,$J = \gamma /4$\\
\end{tabular}
\caption{Summary of the mean-field solutions for a single TASEP with fixed entry and exit rate.  These relations become exact in the $L\to \infty$ limit. Note that here $\alpha$ and $\beta$ denote the dimensionless parameters as defined in the text.\label{eq::TASEP} }
\end{table}

We begin our discussion by recalling the behavior of a single TASEP with fixed hopping rate $\gamma$, entry rate $\tilde\alpha$ and exit rate $\tilde\beta$ \cite{schmittman_statistical_1995,derrida_exactly_1998,schutz_exactly_2001,blythe_nonequilibrium_2007,chou_non-equilibrium_2011}. For the purposes of this paper we discuss only the mean-field results for the steady-state average particle density (number of particles per unit length) and current on the lattice, which become exact in the limit of a large lattice.
We first define rescaled dimensionless entry and exit parameters $\alpha := \tilde\alpha/\gamma$ and  $\beta := \tilde\beta/\gamma$ (i.e. we use the hopping rate $\gamma$ to define the unit of time). 
The (mean-field) density $\rho$ of particles on the lattice takes one of three values, depending on $\alpha$ and $\beta$, as summarized in Table \ref{eq::TASEP}.  The current $J$ is related to the density by $J=\gamma \rho(1-\rho)$. In the low density (LD) phase, where $\alpha<\beta$ and $\alpha< 1/2$, the behavior of the system is dictated by the entry rate and $\rho=\alpha$. In the high density (HD) phase, where $\alpha>\beta$ and $\beta< 1/2$, the exit rate is limiting, particles queue on the lattice, and $\rho =  1-\beta$. When both entry and exit rates are large, $\alpha,\beta\geqslant 1/2$, neither rate is limiting and the system is in the maximal current (MC) phase, for which $\rho=1/2$. The transition from either LD or HD to the MC phase is continuous in the density, but the transition between  the  LD and HD phases is discontinuous in the density. 
On the HD-LD transition line, where $\alpha=\beta$ and $\alpha, \beta < 1/2$, there is ``coexistence'' between the LD and HD phases: part of the lattice takes on the density of the LD phase while the remaining part takes on the density of the HD phase, with a well-defined boundary between these domains, called \emph{shock}. This scenario is known as a shock phase (SP) \cite{derrida_exactly_1998}. 

\begin{figure}[htbp]
	\centering
	\includegraphics[width=0.45\textwidth]{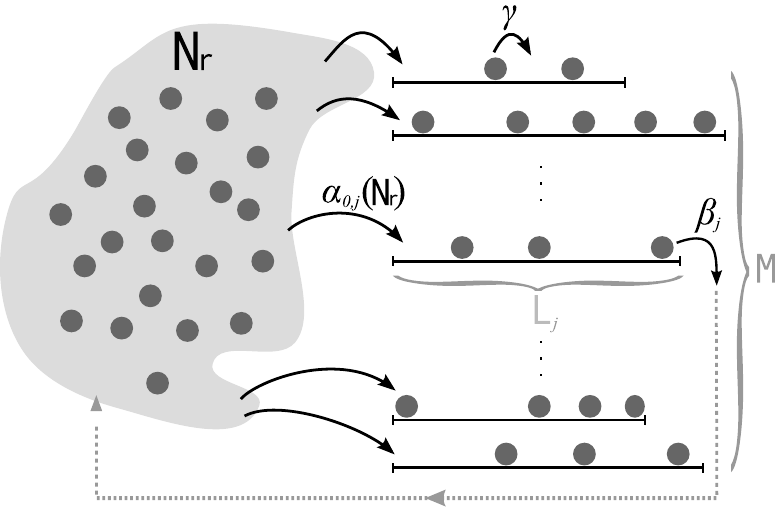}
	\caption{Schematic illustration of the multi-track TASEP with a finite pool of particles. Different lattices may have different lengths, entry and exit rates. The entry rate of particles onto a given lattice depends linearly on the concentration $N_r/V$ of free particles, where $N_r$ is the number of particles in the reservoir, and $V$ is the volume of the reservoir (shaded region). \label{fig::1}}
\end{figure}

In this paper, we consider the case shown in Figure \ref{fig::1}, in which multiple lattices with different lengths, entry and exit parameters compete for a finite number of particles.  The \emph{reservoir} of free particles is assumed to have a finite volume $V$~\footnote{This scenario is typical of biochemical situations (e.g. motor proteins on cytoskeletal filaments) in which both particles and tracks are confined in a finite volume, and the rate of particle binding events per track is proportional to the free particle concentration}. When a particle exits a lattice, it immediately enters  the reservoir, from where it is free to enter any other (or the same) lattice. We assume that free particles are well-mixed (not correlated) and homogeneously distributed within the reservoir. Since particles are conserved, the total number of particles $N$, can be written as 
\begin{equation}
	\label{eq::N}
	N = N_r + N_l \;,
\end{equation}
where $N_r$ denotes the number of particles in the reservoir and $N_l$ the number of particles on the lattices. In turn, $N_l$ can be found by summing over the lattices:
\begin{equation}\label{eq:1}
N_l = \sum_{j=1}^M \rho_j L_j \;,
\end{equation}
where $M$ is the total number of lattices, $\rho_j$ is the density of particles on lattice $j$ and $L_j$ is the length of lattice $j$. 

The entry rate of particles onto any given lattice depends on the density $N_r/V$ of particles in the reservoir. Since the reservoir is well-mixed, we expect this rate to be linear in the particle density, so that an individual TASEP $j$ experiences an injection rate $\alpha_j$ given by 
\begin{equation}
	\label{alpha}
	\alpha_j : = \alpha_{0,j} \frac{N_r}{V}\;,
\end{equation}
where $\alpha_{0,j}$ is the ``intrinsic'' affinity of lattice $j$ for particles (with units of volume/time). 
This relation implies that the rate of particle binding events per lattice is proportional to the free particle concentration: this is expected to be true as long as the 
reservoir does not become too crowded with particles.

In the limit of a large total number $N$ of particles ($N \to \infty$ at constant $N/V$),  $N_r = N - N_l \gg N_l$, so that $N_r/V$ will tend to the total particle concentration $N/V$, and we recover the standard TASEP with constant entry and exit rates. Since the maximum value of $N_l$ is $\sum_j L_j = L M$ (denoting $L := \langle L_j \rangle$ as the average over all lattices), the standard TASEP limit is approached when $N \gg LM$. This means that the finite reservoir plays an important role only for $N/LM \sim 1$ or smaller. Following previous work \cite{adams_far--equilibrium_2008,cook_feedback_2009,cook_competition_2009}, the exit rate from lattice $j$ is assumed to be independent of the reservoir, being simply given by its ``intrinsic'' exit rate $\beta_j$.

Within the mixed population of TASEPs, the phase behavior of a given lattice $j$ is completely determined by its ($N_r$-dependent) effective entry rate $\alpha_j$ and exit rate $\beta_j$, according to the rules for a single TASEP, as listed in Table \ref{eq::TASEP}. For any given value of the reservoir particle number $N_r$ (assuming fixed $V$), subpopulations of lattices will be in the LD, HD and MC phases, depending on their values of $\alpha_j$ and $\beta_j$~\footnote{In this work, we use a different convention from previous authors~\cite{adams_far--equilibrium_2008, cook_feedback_2009, cook_competition_2009} to define the phases of the single-  and multi-track TASEP with finite resources. These authors use a saturating function for the injection rate $\alpha$ and label the phase of the system according to the phase that would be obtained in the infinite reservoir limit (i.e. at the saturation value of the injection rate). We instead define the phase of the lattices as a {\em{reservoir density-dependent}} variable, determined by the value $\alpha(N_r/V)$ of the injection rate.}. We can therefore write the total number of particles on the lattices as 
\begin{equation}\label{eq:2a}
N_l = N_{LD} + N_{HD} + N_{MC} \;,
\end{equation}
where $N_{LD}$, $N_{HD}$ and $N_{MC}$ are the total numbers of particles on lattices in the LD, HD and MC phases, respectively. The density of particles on each lattice depends on which phase it is in, according to the relations in Table \ref{eq::TASEP}. Equation~(\ref{eq:2a}) can therefore be rewritten as
\begin{equation}\label{eq:2}
N_l(N_r) =  \sum_{j, LD} \alpha_j(N_r) L_j + \sum_{j, HD} (1-\beta_j) L_j + \sum_{j, MC}  L_j/2 \;.
\end{equation}
Here, we explicitly note that $N_{LD}$ depends on the reservoir particle number $N_r$ through the dependence of the effective injection rate $\alpha_j$ on $N_r$. The three sums in  Eq.~(\ref{eq:2}) are over lattices in the LD, HD and MC phases: because $\alpha_j$ depends on $N_r$, changes in the reservoir particle number will drive phase transitions on the lattices, so that these three sums will encompass different lattice subpopulations for different values of $N_r$. We can then combine Eqs.~(\ref{eq::N}) and (\ref{eq:2}) to express the relation between the reservoir particle number $N_r$ and the total particle number $N$:
\begin{equation}\label{eq:3}
N = N_r + \sum_{j, LD} \alpha_j(N_r) L_j + \sum_{j, HD} (1-\beta_j) L_j + \sum_{j, MC}  L_j/2 \;.
\end{equation}
Equation~(\ref{eq:3}) forms the basis of our theoretical approach; in the remainder of the paper we explore its implications, first for some simple cases and then for more complex scenarios.

\section{A homogenous population of TASEP\lowercase{s} coupled to a finite reservoir}
\label{sec::population}

We first consider the case of a homogeneous population of lattices. Our system thus contains $M$ identical lattices of length $L$, intrinsic injection rate $\alpha_0$ and exit rate $\beta$, coupled to a finite pool of $N$ particles. Since the lattices are identical, in our mean-field approach each lattice experiences the same effective injection rate $\alpha = \alpha_0 N_r/V$. This implies that all the lattices are in the same phase; we refer to this as the  ``global phase'' of the system. The relation between $N$ and $N_r$ is now given by:
\begin{equation}
	N = 
	\begin{cases}
	N_r + L M\alpha_0 N_r/V & \alpha_0 N_r/V < \beta, \,\alpha_0 N_r/V < 1/2\\
	
	N_r +LM(1-\beta) & \alpha_0 N_r/V > \beta, \,\beta < 1/2\\
	N_r + LM/2 & \alpha_0 N_r/V \geqslant 1/2,\, \beta \geqslant 1/2\;,
	\end{cases}
	\label{eq::r2}
\end{equation}
where the first, second and third relations refer to the global LD, HD and MC phases, respectively.

\subsection{Single TASEP}
\label{sec:single}

For the purposes of illustration, we first consider the case of a single ($M=1$) TASEP of length $L$. For this test case, our approach corresponds closely to the calculation presented for a saturating function $\alpha(N_r)$ by Adams {\em{et al}} \cite{adams_far--equilibrium_2008}. The relation between the total particle number and the number of particles in the reservoir is given by Eqs.~(\ref{eq::r2}), taking $M=1$.
\begin{figure}[htbp]
	\centering
	\vspace{0cm}\includegraphics[width=0.5\textwidth]{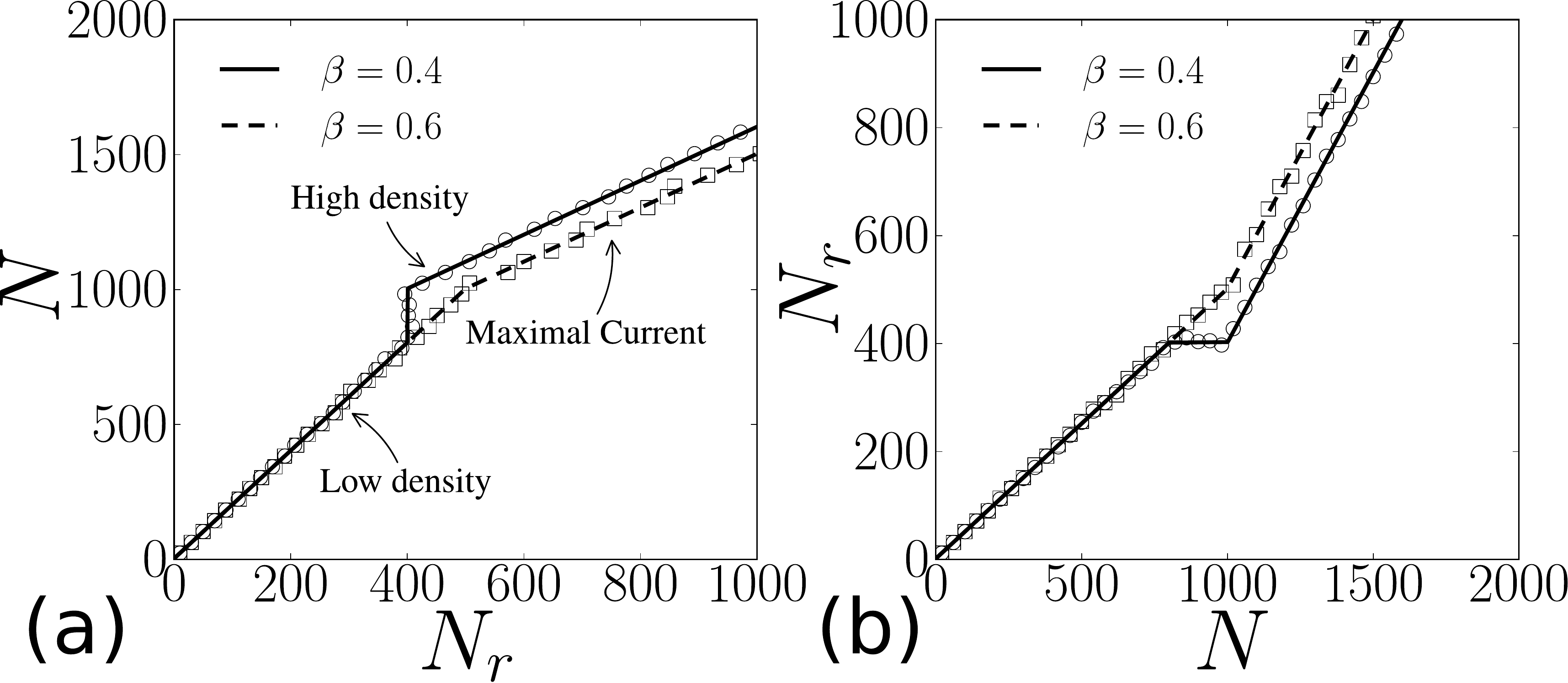}
	\caption{(a) Relation between total particle number $N$ and reservoir particle number $N_r$ for a single TASEP with a finite reservoir of particles, from Eqs.~(\ref{eq::r2}) with $M=1$, $L=1000$ and $\alpha_0/V = 0.001$. The solid and dashed lines show the mean-field mapping, Eqs.~(\ref{eq::r2})  for $\beta = 0.4$ and $0.6$ respectively, while the circles and squares show the results of kinetic Monte Carlo simulations for the same parameter sets. (b) The same data as in (a) but inverted, i.e., showing the reservoir particle number $N_r$ as a function of the total particle number $N$. \label{fig::single}}
\end{figure}

Figure~\ref{fig::single}a (solid and dashed lines) illustrates the mathematical mapping between $N_r$ and $N$ given by Eqs.~(\ref{eq::r2}).  If $\beta<1/2$, then when $\alpha=\beta$ the lattice enters the HD phase and the number of particles on the lattice jumps discontinuously. The results of the mean-field theory are in very good agreement with continuous time kinetic Monte Carlo simulations for a single TASEP coupled to a finite reservoir of particles, for the same parameter sets (circles and squares). 
 
In most realistic scenarios, however, we expect that the  total number of particles $N$ is fixed, while the reservoir number $N_r$ achieves a self-adjusting steady-state value depending on $N$ and the parameters of the lattices. We therefore need to invert the mapping of  Eqs.~(\ref{eq::r2}) to find $N_r(N)$: the number of particles in the reservoir for a fixed total particle number. For a single lattice, this inversion is shown in  Fig. ~\ref{fig::single}b. For $\beta<1/2$ the LD-HD phase transition occurs over a finite range of $N$; i.e. there exists  a  range of values of $N$ for which the number of particles in the reservoir $N_r$ is independent of $N$. In this range of values of $N$, the lattice is in the shock phase, with HD and LD domains. If a new particle is added to the system it is quickly absorbed by the lattice rather than staying in the reservoir, increasing the size of the HD domain. Likewise, if a particle is removed from the system, the relative size of the HD and LD domains will adjust to keep the number of particles in the reservoir constant. Thus, while the lattice is in the shock phase, $N_r$ is independent of $N$: the shock phase lattice buffers the particle reservoir to fluctuations in the total particle number. It is interesting to note that our mean-field theory captures this key qualitative feature of the system, even though it deals only with the average density of particles on the lattice and cannot  resolve the dynamics of the LD-HD ``domain wall''.

\subsection{Multiple TASEPs}
\label{sec::homo}
Building on our results for a single TASEP, we next consider a system consisting of M identical lattices. This system shows qualitatively similar behavior to that illustrated in Fig.~\ref{fig::single} for the single TASEP case: if $\beta < 1/2$, the system undergoes a global LD-HD transition when $\alpha_0 N_r/V =\beta$, and this produces a range of values of $N$ for which the particle reservoir number $N_r$ is independent of $N$. The width of this plateau in $N_r(N)$ is exactly the height of the step in $N(N_r)$ and can be easily obtained from Eqs. (\ref{eq::r2}): substituting the critical condition $N_r =\beta V/\alpha_0$ and taking the difference between line 2 and line 1 in Eqs. (\ref{eq::r2}) gives the plateau width 
\begin{equation}
\label{eq::8}
\Delta N = LM(1-2\beta)   \; . 
\end{equation}
Hence, $\Delta N$ increases linearly with the number of lattices $M$.

\begin{figure}[htbp]
	\centering
	\includegraphics[width=0.47\textwidth]{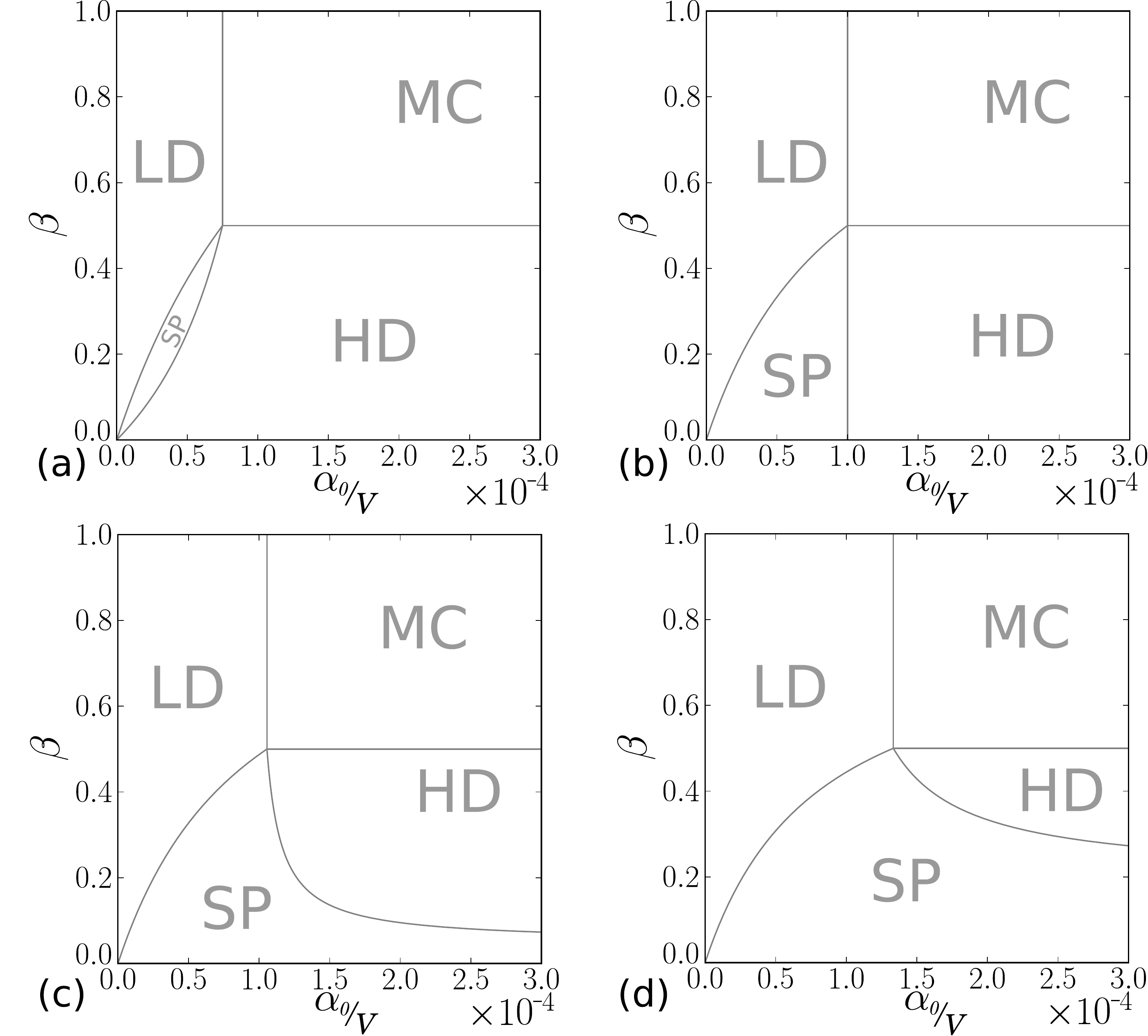}
	\caption{Phase diagrams in the $(\alpha_0 /V, \beta)$ plane of the multi-track TASEP for a fixed set of parameters ($N$,$L$,$M$). (a) $N=1.5\cdot 10^4$, (b) $N=L M =10^4 $, (c) $N=9.5\cdot 10^3$, (d) $N=8\cdot 10^3 $. In all cases $L=10^3$ and $M=10$. These choices of parameters represent, as discussed in Sec.~\ref{sec:theory}, high (a), medium (b-c) and low (d) levels of availability of free particles. Note that in this figure we consider only cases in which $2N>LM$. If instead $2N<LM$, there are not enough particles to bring all lattices into the MC phase and the phase diagram contains only the LD and SP phases.   \label{fig::2}}
\end{figure}

We can also use Eqs.~(\ref{eq::r2}) to obtain a global phase diagram for the system in the parameter space of the intrinsic entry and exit rates ($\alpha_0/V,\beta$), for fixed $N, L$ and $M$, by substituting the critical condition $N_r =\beta V/\alpha_0$ that defines the phase boundary into the relations for $N(N_r)$ in Eqs.~(\ref{eq::r2}) ~\footnote{For example, the phase transition from LD to SP can be obtained by substituting the critical value $N_r = \beta V/\alpha_0$ in the first line of Eq. (\ref{eq::r2}). This gives an implicit form for the phase transitions in the parameters $\{\alpha_0,\beta,N,L,M,V\}$. Fixing $N$, $M$, $L$ and $V$, and solving for $\beta$, the equation gives the the explicit form of the phase transition line. All other phase transitions can be found in the same way by substituting $N_r$, as shown in the Appendix. We note that in the parameter space $(\alpha, \beta)$ the phase diagram would trivially be that of a single TASEP with infinite resources (fixed $\alpha$ and $\beta$).}. Figure~\ref{fig::2} shows this phase diagram, for several different values of the total number of particles $N$. Interestingly, as {\bf{$N$}} decreases  (Figures~\ref{fig::2}a-d), a finite region of the phase diagram emerges where the lattices are in the shock phase (SP): here, the particle number is too high for the LD phase but insufficient to allow the TASEPs to reach the HD phase. As discussed above, in this SP region of the phase diagram, the reservoir particle number $N_r$ remains constant such that $\alpha_0 N_r/V=\beta$.
The extent of this SP region decreases as the number of particles in the reservoir increases, and for  $N\gg LM$ the phase diagram tends to that of a single TASEP with infinite resources and $\alpha = \alpha_0 N/V$. 

\begin{figure}[htbp]
	\centering
	\includegraphics[width=0.5\textwidth]{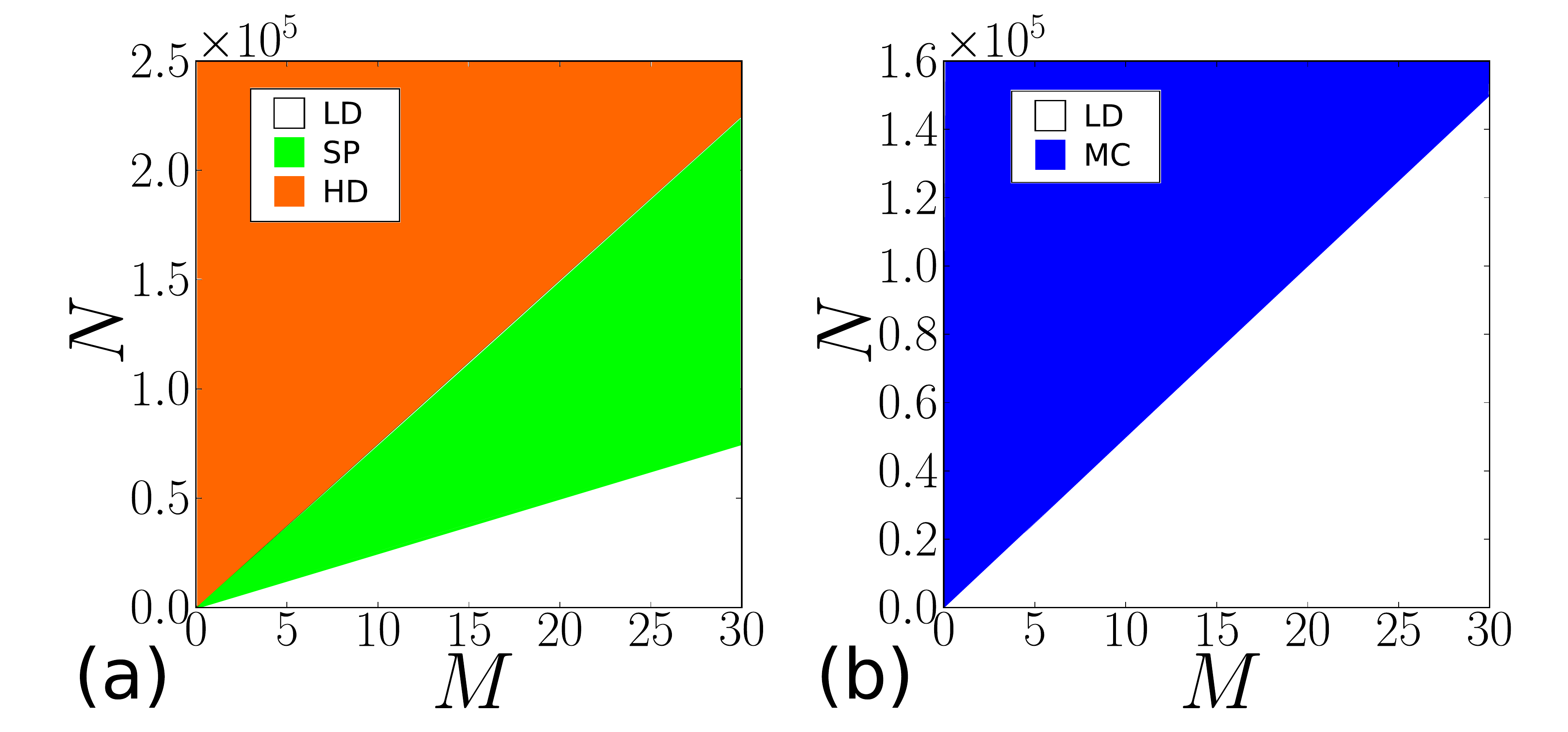}
	\caption{Multi-track TASEP phase diagram in the parameter space  ($N$, $M$), for $\alpha_0/V=5\cdot 10^{-3}$ and $L=10^4$. Panel (a) shows a case with $\beta=0.25$, while in (b) we show a case in which the exit rate is not limiting, $\beta \geqslant 1/2$ (note that in this case the boundary between LD and MC does not depend on the precise value of $\beta$). \label{fig::3}}
\end{figure}

While Fig.~\ref{fig::2} provides insight into the physics of the system, in a real experimental situation we expect that the control parameters are  most likely to be the number of particles $N$ and the number of lattices $M$, rather than $\alpha_0$ and $\beta$. It is therefore useful to plot phase diagrams in the parameter space $(N, M)$, for fixed $\alpha_0/V$ and $\beta$. This can be achieved by reformatting the boundary conditions for the global LD, HD and MC phases in terms of $N$ and $M$, using the relations listed in Eqs.~(\ref{eq::r2}). For example, for the system to be in the global LD  phase we require $\alpha_0 N_r/V < \beta $ and $\alpha_0 N_r/V < 1/2$. Using the relation $N = N_r + L M\alpha_0 N_r/V$ which holds for the LD phase, the first inequality can be reformulated as  $N < (V \beta / \alpha_0 + LM\beta)$ while the second becomes $N < (V/ \alpha_0 + LM)/2$. This procedure allows us to build phase diagrams in the  $(N, M)$ parameter space. Figure ~\ref{fig::3}a shows a case where $\beta < 1/2$, so that the phase diagram contains a shock phase region, while Figure ~\ref{fig::3}b shows a case where $\beta \geqslant 1/2$, so that the system instead makes a continuous transition between the LD and MC phases. Because of competition among the lattices for particles, the global state of the system depends on both $N$ and $M$. Increasing the number of lattices $M$ decreases the number of particles per lattice, pushing the system towards the LD phase, while increasing the number of particles $N$ increases the particles per lattice,  pushing the system towards the HD or MC phase. Table ~\ref{tab::conditions} in Appendix \ref{app::homo} summarizes the boundaries among the global phases of this multi-track TASEP, in both parameter spaces $(N, M)$ and $(\alpha_0/V,\beta)$.\\

\section{A mixed population of TASEP\lowercase{s}}\label{sec::mixed}
We now move on to the more relevant case, where the lattices are not all identical. Here, we find that the coupling between lattices induced by the finite reservoir has interesting and non-trivial effects. We first consider a population  composed of two different types of lattices: $M^{(1)}$ lattices with intrinsic injection rate $\alpha_{0}^{(1)}$ and $M^{(2)}$ lattices with intrinsic injection rate $\alpha_{0}^{(2)} > \alpha_{0}^{(1)}$. Note that here we introduce a new notation: we use upper indexes in brackets (e.g., $\alpha_0^{(i)}$) to indicate properties shared by all lattices in the same subpopulation, as opposed to  lower indexes (e.g., $\alpha_{0,j}$) which we used to denote the properties of individual lattices.  For the sake of simplicity we suppose that all the lattices have the same length $L$ and exit rate $\beta$. Our methodology can easily be extended to the case of different $L$ and $\beta$ (see Section \ref{sec::arbitrary}). 

Because the two lattice subpopulations have different values of $\alpha_0$, they will undergo the LD-HD or LD-MC phase transition at different values of the reservoir particle number $N_r$. 
For the same value of $N_r$, the two subpopulations of lattices can therefore be in different phases. The total number of particles can be expressed in terms of the particle densities  $\rho^{(1)}$ and $\rho^{(2)}$ on the type-1 and type-2 lattices:
\begin{equation}
	N = N_r + \rho^{(1)}LM^{(1)} + \rho^{(2)}LM^{(2)} \;,
	\label{eq::r_two_types}
\end{equation}
The densities  $\rho^{(i)}$ ($i=1,2$) depend on the phase of the lattice: if $\beta<1/2$ one has
\begin{equation}
	\rho^{(i)} = 
	\begin{cases}
	\cfrac{\alpha_{0}^{(i)}}{V}  N_r & \text{if } \alpha_{0}^{(i)} N_r/V < \beta  \qquad(LD) \\ 
	1-\beta & \text{if } \alpha_{0}^{(i)} N_r/V > \beta  \qquad(HD) \;,
	\end{cases}
	\label{eq::densities1}
\end{equation}
while if $\beta \geqslant 1/2$ the densities are
\begin{equation}
	\rho^{(i)} = 
	\begin{cases}
	\cfrac{\alpha_{0}^{(i)}}{V}  N_r & \text{if } \alpha_{0}^{(i)} N_r/V < 1/2  \qquad(LD) \\ 
	1/2 & \text{if }\alpha _{0}^{(i)} N_r/V \geqslant 1/2  \qquad(MC) \;.
	\end{cases}
	\label{eq::densities2}
\end{equation}
We can express Eq.~(\ref{eq::r_two_types}) in a compact way by making use of the  Heaviside function (defined as $\theta(z)=1$ for $z \geqslant0$ and $\theta(z)=0$ otherwise). For $\beta<1/2$ this results in:
\begin{eqnarray}
	N  &=& N_r\\ \nonumber
	  + & \!\! LM^{(1)}&\!\!\!\! \left[\frac{\alpha_{0}^{(1)} N_r}{V} \theta \! \left(\beta - \frac{\alpha_{0}^{(1)}N_r}{V}\right) \!\! + (1-\beta)\theta \! \left(\frac{\alpha_{0}^{(1)}N_r}{V}-\beta\right)\right] \\ \nonumber
	 + &  \!\! LM^{(2)}&\!\!\!\! \left[\frac{\alpha_{0}^{(2)} N_r}{V} \theta \! \left(\beta - \frac{\alpha_{0}^{(2)}N_r}{V}\right) \!\! + (1-\beta)\theta \! \left(\frac{\alpha_{0}^{(2)}N_r}{V}-\beta\right)\right],
\end{eqnarray}
which can be rearranged to give:
\begin{eqnarray}
	\label{eq::r_two_types_theta}
N &=& N_r + \cfrac{\alpha_{0}^{(1)}N_r}{V} LM^{(1)} + \cfrac{\alpha_{0}^{(2)}N_r}{V} LM^{(2)} \\ \nonumber
	& &+ L M^{(1)} (1-\beta -\cfrac{\alpha_{0}^{(1)}N_r}{V}) \; \theta\left(N_r-\cfrac{V\beta}{\alpha_{0}^{(1)}}\right) \\ \nonumber
	& &+  L M^{(2)} (1-\beta-\cfrac{\alpha_{0}^{(2)}N_r}{V})\; \theta\left(N_r-\cfrac{V\beta}{\alpha_{0}^{(2)}}\right)\;.
\end{eqnarray}
For the case where $\beta>1/2$ we could write an equivalent equation, using instead the densities and boundary conditions appropriate for the LD to MC phase transition. It turns out however that this is exactly equivalent to Eq.~(\ref{eq::r_two_types_theta}), but with $\beta$ replaced by $1/2$. Equation~(\ref{eq::r_two_types_theta}) can therefore be used to describe the full behavior of the system, for any value of $\beta$, with the proviso that for $\beta>1/2$, we simply set $\beta=1/2$ in the equation.
\begin{figure}[htbp]
	\centering
	\includegraphics[width=0.4\textwidth]{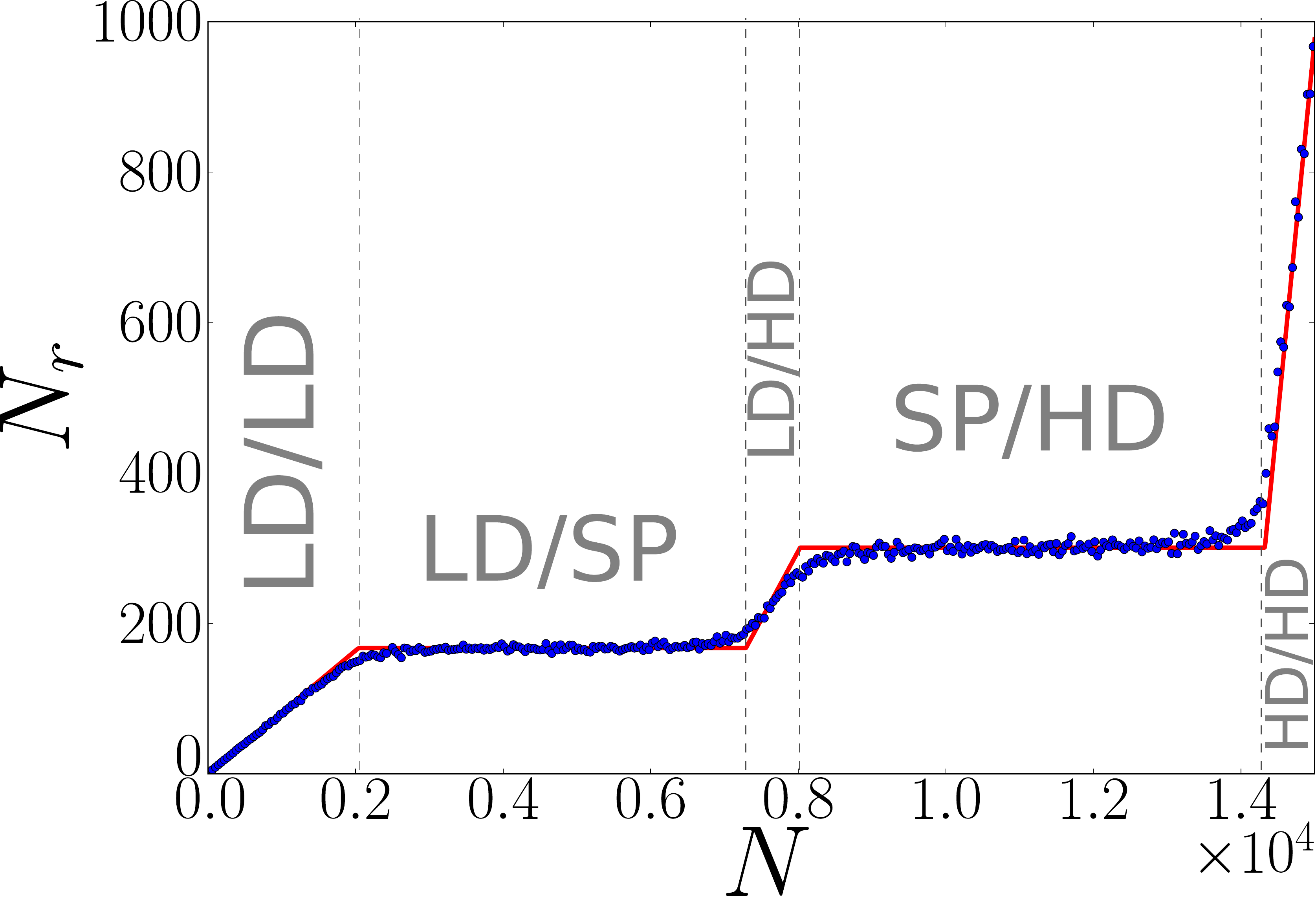}
\caption{Reservoir particle number $N_r$ as a function of total particle number $N$ for the system with two subpopulations of lattices  (color online), with $\alpha_0^{(1)}/V=5 \cdot 10^{-4}$, $\alpha_0^{(2)}/V=9 \cdot 10^{-4}$, $\beta=0.15$, $L=300$, $M^{(1)}=30$, and $M^{(2)}=25$.  The full red line shows the results of the mean-field theory, produced by numerical inversion of Eq.~(\ref{eq::r_two_types_theta}); the blue circles are the results of kinetic Monte Carlo simulations with the same parameter set. For the mean-field theory, the first derivative of $N(N_r)$ is discontinuous at the critical points given by Eqs.~(\ref{eq::crit_points1}).  The width $\Delta N$ of the first plateau is given by $\Delta N = LM^{(2)} (1-2\beta)$ while the second plateau has $\Delta N = LM^{(1)} (1-2\beta)$, see Eq.~(\ref{eq::8}). The phases of the two lattices subpopulations are also labelled: for example, LD/HD denotes a regime with the type-1 lattices in the LD phase and the type-2 lattices in the HD phase. \label{fig::4}}
\end{figure}

From Equation~(\ref{eq::r_two_types_theta}) we can obtain the physical relevant relation $N_r(N)$ by inversion (as in Sec. III B). $N_r(N)$ is plotted in Figure 5, for the case $\beta<1/2$: the two $\theta$-steps of Equation~(\ref{eq::r_two_types_theta}) appear as two plateaus, corresponding to two regimes where the reservoir is buffered due to the emergence of a shock phase (SP) on one lattice subtype. In the case $\beta \geqslant 1/2$ we do not obtain this effect, because the particle density on a lattice is continuous across the LD to MC transition. Figure~\ref{fig::4} also shows the results of kinetic Monte Carlo simulations for a multi-TASEP system coupled to a finite reservoir, with the same parameter set. The agreement between the mean-field solution and the simulations is very good; the slight discrepancy around the phase transitions is due to finite size effects (the mean-field solution becomes exact only in the limit $L \to \infty$).

Equation~(\ref{eq::r_two_types_theta}) also provides a simple way to determine the phase boundaries for this system; these occur as the discontinuities of the Heaviside step functions are approached from above and below -- {\em{i.e.}} at $N_r  \to \left(V\beta/\alpha_{0}^{(2)}\right)^\pm$ and $N_r  \to \left(V\beta/\alpha_{0}^{(1)}\right)^\pm$. Appendix \ref{app::cirit_points} gives explicit forms for these phase boundaries.

\begin{figure}[htbp]
	\centering
	\includegraphics[width=0.48\textwidth]{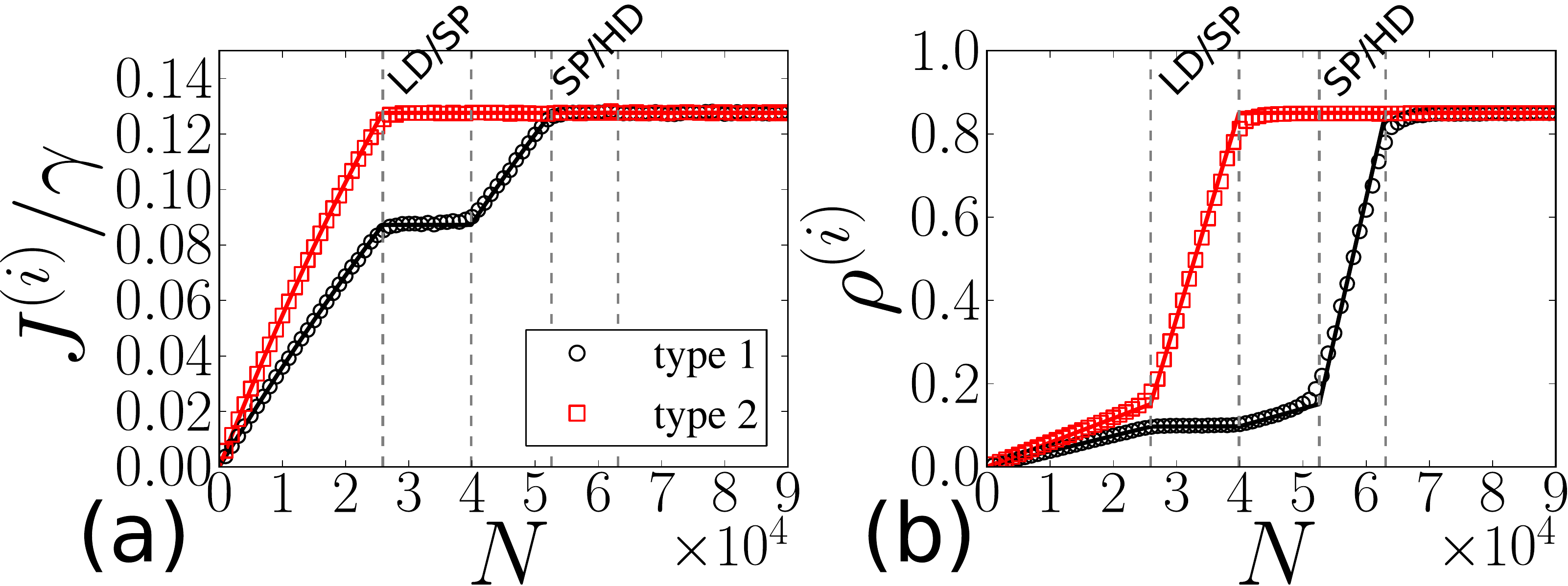}
	\caption{Current per lattice $J^{(i)}$ (a) and density $\rho^{(i)}$ (b) on type-1 and type-2 lattices for the system with two lattice subpopulations. The red and black lines represent the mean-field theory results for the type-1 and type-2 subpopulations of lattices respectively (color online); the squares and circles show kinetic Monte Carlo simulation results. The parameters used were  $\alpha_0^{(1)}/V=4.5 \cdot 10^{-6}$, $\alpha_0^{(2)}/V=7 \cdot 10^{-6}$, $\beta=0.15$, $L=1000$, $M^{(1)}=15$, and $M^{(2)}=20$. Note that the units of current are the particle hopping rate $\gamma$. \label{fig::5}}
\end{figure}

Figure \ref{fig::5}  shows that the current per lattice and the particle density on the individual lattices exhibit a remarkable dependence on the total particle number $N$. For the range of values of $N$ over which lattice type 2 is in the SP,  the current and density on the lattices of type 1 remain constant, even though these lattices are far away from a phase transition. This buffering effect occurs because the  reservoir particle number remains constant while any lattice subtype is in the SP; this fixes $\alpha$, and hence the current and density, of all lattices which are in the LD-phase (see Table 1). Hence, lattices of type 1 are affected by the phase transition on lattice type 2, through the coupling to the finite particle reservoir. Note that while lattice type 2 is in the SP, its current remains constant, but its density increases linearly with $N$. This reflects the fact that, in the ``buffering regime'', the SP lattices absorb particles as $N$ increases, keeping  $N_r$ constant. The same observation applies to lattice type 1 in the second buffering region.

\begin{figure}[htbp]
	\centering
	\includegraphics[width=0.5\textwidth]{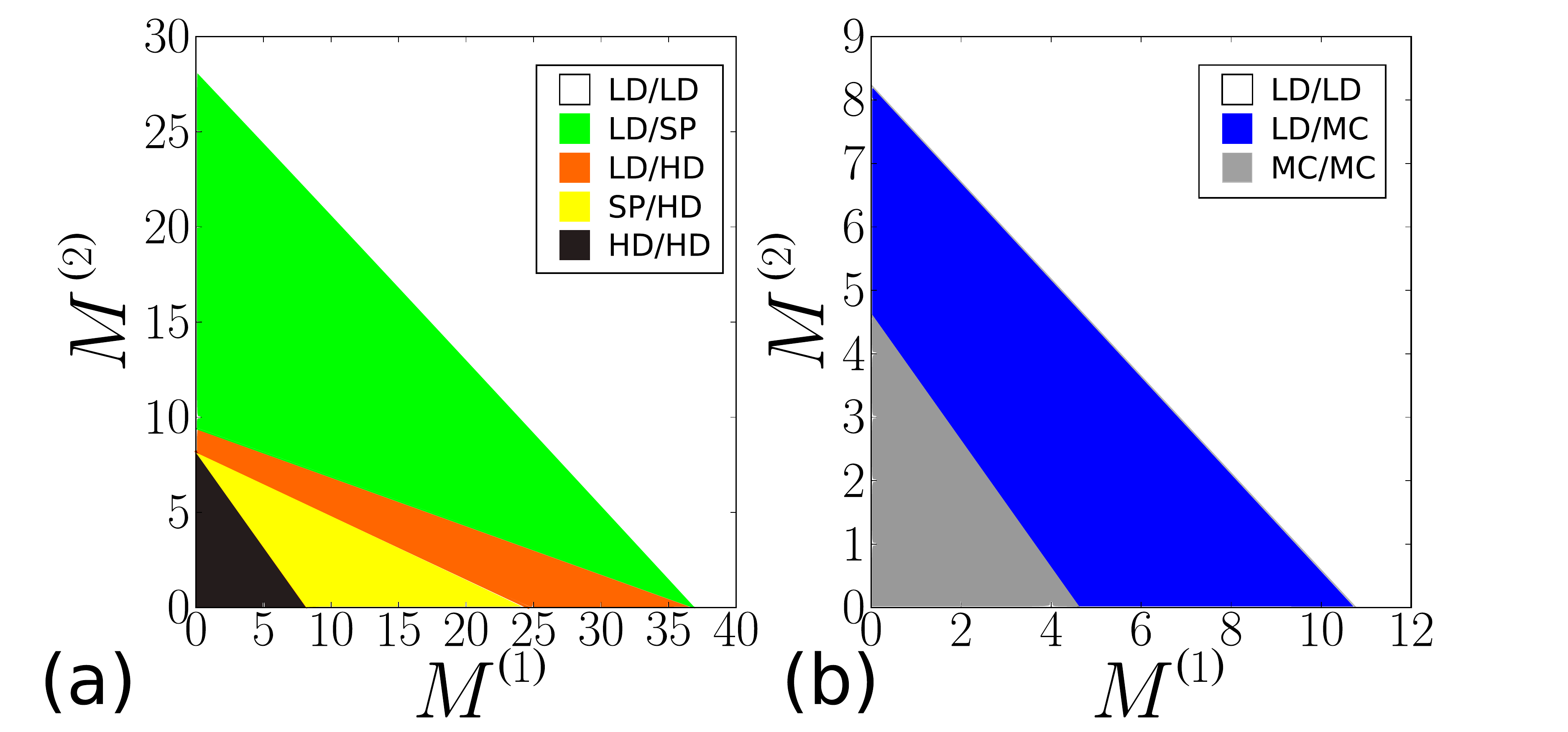}
	\caption{Multi-track TASEP phase diagram in the parameter space  ($M^{(1)}$, $M^{(2)}$). Panel (a) presents a case with $\beta=0.25$, while in (b) we show the situation in which the depletion rate is not limiting, $\beta \geqslant 1/2$. The other parameters used are $\alpha_1/V=6.5\cdot10^{-3}$, $\alpha_2/V=8.5\cdot10^{-3}$, $L=10^3$, $N=10^4$. \label{fig::6}}
\end{figure}

To conclude this section, we investigate the phase behavior of this system as a function of the numbers $M^{(1)}$ and $M^{(2)}$ of lattices in the two subpopulations. The parameter space  ($M^{(1)}$, $M^{(2)}$) is likely to be relevant experimentally: for example, in intracellular transport problems,  $M^{(1)}$ and $M^{(2)}$ might represent the number of cytoskeletal filaments of two different types along which motor proteins can travel. Figure~\ref{fig::6} shows the phase diagram of the system in the ($M^{(1)}$, $M^{(2)}$) plane for (a) $\beta<1/2$ and (b) $\beta \geqslant 1/2$. It is clear that the competition for particles plays an important role: the phase behavior of each lattice subpopulation depends strongly on the size of the other subpopulation.

\section{A mixed population with arbitrary distribution of boundary rates}
\label{sec::arbitrary}
We now formulate the concepts of Sections~\ref{sec:theory}-\ref{sec::mixed} in a general way, to allow us to describe a mixed lattice population with an arbitrary 
distribution of parameter values.
As in the preceding discussion -- see Eqs.~(\ref{eq::N}) and (\ref{eq:3}) -- our strategy is to express the total number of particles $N = N_r + N_{LD} +N_{HD}+N_{MC}$ as a function of the boundary parameters of the lattices.

The total number of lattices is denoted by $M$ and the normalized distribution of boundary rates and lengths (i.e. their relative frequencies) for the mixed lattice population is denoted by $P(\alpha_0,\beta,L)$. 
Such a continuous distribution could be appropriate in situations where lattice parameter values are sensitive to small environmental changes, or where the total number of lattices is very large.
We can then express the total number of particles on lattices in the LD, HD and MC phases as:
\begin{eqnarray}
\label{eq::integrals}
 \nonumber N_{LD} &=& M\int_0^{\infty}dL\iint\limits_{\text{LD}} L \frac{\alpha_0 N_r}{V}  \,P(\alpha_0,\beta,L) \,d\alpha_0\, d\beta\\ \nonumber
N_{HD} &=& M\int_0^{\infty}dL\iint\limits_{\text{HD}} L \left(1-\beta\right) \,P(\alpha_0,\beta,L) \, d\alpha_0\, d\beta  \\ 
N_{MC} &=& M\int_0^{\infty}dL\iint\limits_{\text{MC}} \frac{L}{2} \,P(\alpha_0,\beta,L)  d\alpha_0 \,d\beta \;,
\end{eqnarray}
where the integration limits are determined by the constraints on $\alpha=\alpha_0 N_r/V$ and $\beta$ for the LD, HD and MC regions, respectively, as defined in Table \ref{eq::TASEP}. Note that the MC case can be recovered as a special case of the HD with $\beta=1/2$. Therefore, we can limit our analysis to $\beta \leqslant 1/2$, i.e. assuming $P(\beta>1/2)=0$~\footnote{We can restrict our study to the LD and HD phases by redefining the distribution of parameters as follows: $P_{new}(\alpha_o, \beta=1/2,L) := \int_{1/2}^{\infty} d\beta \; P(\alpha_o, \beta,L)$, $P_{new}(\alpha_o, \beta>1/2,L) := 0$, and $P_{new}(\alpha_o, \beta<1/2,L) := P(\alpha_0,\beta<1/2,L)$. From now on, for the sake of simplicity, in the text we refer to $P_{new}(\alpha_o,\beta,L)$ as to just $P(\alpha_o,\beta,L)$.}.

We also assume that the  lengths of the lattices are independent of the boundary rates, such that $P(\alpha_0,\beta,L)=P(\alpha_0,\beta)P(L)$. This allows us to perform the integration  over $L$ in Eq.~(\ref{eq::integrals}), yielding  
\begin{eqnarray}
\label{eq::integrals2}
 \nonumber N_{LD} &=& \langle L \rangle M \int_0^{\infty} d\beta \int_0^{\frac{V \beta}{N_r}} d\alpha_0 \frac{\alpha_0 N_r}{V}  \,P(\alpha_0,\beta) \\
N_{HD} &=& \langle L \rangle M  \int_0^{\infty} d\beta \int_{\frac{V\beta}{N_r}}^{\infty} d\alpha_0 \left(1-\beta\right) \,P(\alpha_0,\beta)  \;,
\end{eqnarray}
where  $\langle L \rangle $ is the average lattice length and we have inserted the limits of integration as detailed in Table \ref{eq::TASEP}. Note that since $P(\beta>1/2)=0$ we are free to choose the upper limit of the $\beta$-integration as infinity. This simplifies our further calculations.

After adding and subtracting the term $\langle L \rangle M  \int_0^{\infty} d\beta \int_{V\beta/N_r}^\infty d\alpha_0 \frac{\alpha_0 N_r}{V}  \,P(\alpha_0,\beta)$ and inserting the expressions for the subpopulations into the particle conservation equation, we arrive at the result
\begin{eqnarray}
	\label{r_conserv_2}
	N(N_r)& = & N_r + \langle L \rangle  M \biggl[\frac{N_r\langle \alpha_0 \rangle }{V} \\ \nonumber
	& + &\!\! \int_{0}^{\infty} \!\!\!\! d\beta \int_{V\beta/N_r}^{\infty} \!\!\!\!  d\alpha_0 \, \left(1 - \beta- \frac{\alpha_0 N_r}{V} \right) P(\alpha_0,\beta) \, \biggl] \;,
\end{eqnarray}
where  $\langle{\alpha_0}\rangle = \int_{0}^{\infty} d\beta \int_{0}^{\infty}d\alpha_0 \, \alpha_0 P(\alpha_0,\beta)$ is the average value of $\alpha_0$. Equation~(\ref{r_conserv_2}) is a generalisation of  Eqs.~(\ref{eq:3}) and~(\ref{eq::r_two_types_theta}) to  a continuous distribution of parameters.

\subsection{Discrete distributions}
\label{sec::discrete}
We first consider the case where  our mixed population of lattices contains a finite number of distinct subpopulations with  parameters $(\alpha_0^{(i)},\beta^{(i)})$. In this case  the distribution  $P(\alpha_0,\beta)$ can be expressed as a sum over $\delta$-functions: $P(\alpha_0,\beta)= \sum_i (M^{(i)}/M) \delta(\alpha_{0}^{(i)}-\alpha_0)\delta(\beta^{(i)}-\beta)$, where $M^{(i)}$ is  the number of lattices in subpopulation $i$ and $M$ is the total number of lattices.  Equation~(\ref{r_conserv_2}) then takes the form
\begin{eqnarray}
	\label{r_conserv_discr}
	N(N_r)& = & N_r + \langle L \rangle \left[ M \frac{N_r\langle \alpha_0 \rangle}{V} \right. \\ \nonumber
	&\!\!\! \!\!\! \!\!\! \!\!\!+ &\!\!\! \!\!\!\!\!\! \left. \sum_i M^{(i)} \left(1 - \beta^{(i)} - \frac{\alpha_{0}^{(i)} N_r}{V} \right) \theta\left(N_r - \cfrac{V \beta^{(i)}}{\alpha_{0}^{(i)}}\right) \right],
\end{eqnarray}
which is an extension of Eq.~(\ref{eq::r_two_types_theta}) to an arbitrary number of subpopulations. The function $N(N_r)$ has discontinuities at $N_r= V \beta^{(i)}/\alpha_{0}^{(i)}$, for $\beta^{(i)}<1/2$. For the inverse relation  $N_r(N)$, these become plateaus of width $\Delta N =  \langle L \rangle M^{(i)} \left(1 - 2\beta^{(i)} \right)$. Each lattice type for which $\beta^{(i)}<1/2$ gives rise to a distinct plateau in $N_r(N)$. Within plateau region $i$, lattices of type $i$ are in the shock phase, at the LD-HD phase boundary. Importantly, because the reservoir particle number $N_r$ controls the behavior of the whole system, a plateau in $N_r(N)$ implies that the entry of any subpopulation of lattices into the shock phase is sufficient to make the whole system independent of the total particle number $N$.

Since the relation $N(N_r)$ cannot be inverted analytically, one cannot give a simple prescription for the ranges of values of the total particle number where the system is buffered. However the generic prescription for the regions of $N$ over which the system is independent of $N$ is:
\begin{eqnarray}
\label{PT-prescription}
&&\text{lower boundary of region $i$} : \lim_{N_r \to (V\beta^{(i)}/\alpha_{0}^{(i)})^-} N(N_r)\qquad \,\,\\ \nonumber
&&\text{upper boundary of region $i$} : \lim_{N_r \to (V\beta^{(i)}/\alpha_{0}^{(i)})^+} N(N_r)
\end{eqnarray} 
for all lattice subpopulations $i$ for which $\beta^{(i)}<1/2$. These boundaries occur at the positions of the steps in the $\theta$-functions in Eq.~(\ref{r_conserv_discr}); the upper boundaries occur as the steps are approached from above, and the lower boundaries as the steps are approached from below. It is important to note that  the phase transitions on any given lattice type depend on the parameter values of all the lattices in the system, since $N_r$ is determined by the competition for particles among all the lattices.

\subsection{Continuous distribution}
\label{sec::continuous}

We next consider a scenario where the population of lattices does not contain distinct lattice subtypes, but instead is described by a  continuous probability distribution of lattice parameters $P(\alpha_{0},\beta,L)$. In this case, the conservation equation (\ref{r_conserv_2}) for the particle number does not reduce to a sum of $\delta$-functions, as it does in the discrete case, and consequently there are no discontinuities in the function $N(N_r)$.  As a simple example, let us assume that all the lattices have the same fixed value $\alpha_0=\alpha_0^*$, with a continuous probability distribution for $\beta$: i.e. $P(\alpha_0, \beta) = \delta (\alpha_0-\alpha_0^*)P(\beta)$ (as before we have redefined $P(\beta)$ to consider the MC phase as a special case of the HD phase). The integral over $\alpha_0$ in Eq.~(\ref{r_conserv_2}) then reduces to a step function  $\theta(\alpha_0^*- \beta V/N_r)$, leading to:
\begin{eqnarray}
\label{r_fixed_alpha}
N(N_r) &= &N_r + \langle L \rangle  M \left[ \frac{\alpha_0^* N_r}{V} \right. \\ \nonumber
&+ & \left. \int_0^{\alpha_0^*N_r/V} \left(1-\beta-\frac{\alpha_0^* N_r}{V}\right) P(\beta) \,d\beta \right] \;,
\end{eqnarray}
where the upper limit on the integral reflects the condition $\alpha>\beta$ for the HD phase.

To explore the consequences of Eq.~(\ref{r_fixed_alpha}), we consider the specific case where the distribution of exit rates is Gaussian: $P(\beta) = 1/\sqrt{2\pi \sigma^2} \exp(-(\beta-\langle\beta \rangle)^2/2\sigma^2)$.  In this case, the integral in Eq.~(\ref{r_fixed_alpha}) can be calculated analytically to give:
\begin{eqnarray}
\label{r_rhoc_gauss}
N(N_r) & =& N_r + \frac{ \alpha_0^* N_r}{V} \\ \nonumber
&+& \langle L \rangle  M \left[ 1 - \left( \frac{1}{2}\langle\beta\rangle{\rm erf}((\beta-\langle\beta\rangle)/\sqrt{2}\sigma) \right. \right. \\ \nonumber
&&\left. \left. -  \sqrt{\frac{\sigma^2}{2\pi}} e^{(\beta-\langle\beta\rangle)^2/2\sigma^2}  \right) - \frac{\alpha_0^* N_r}{V}  \right] \;.
\end{eqnarray}
\begin{figure}
\begin{center}
\includegraphics[width=1\columnwidth]{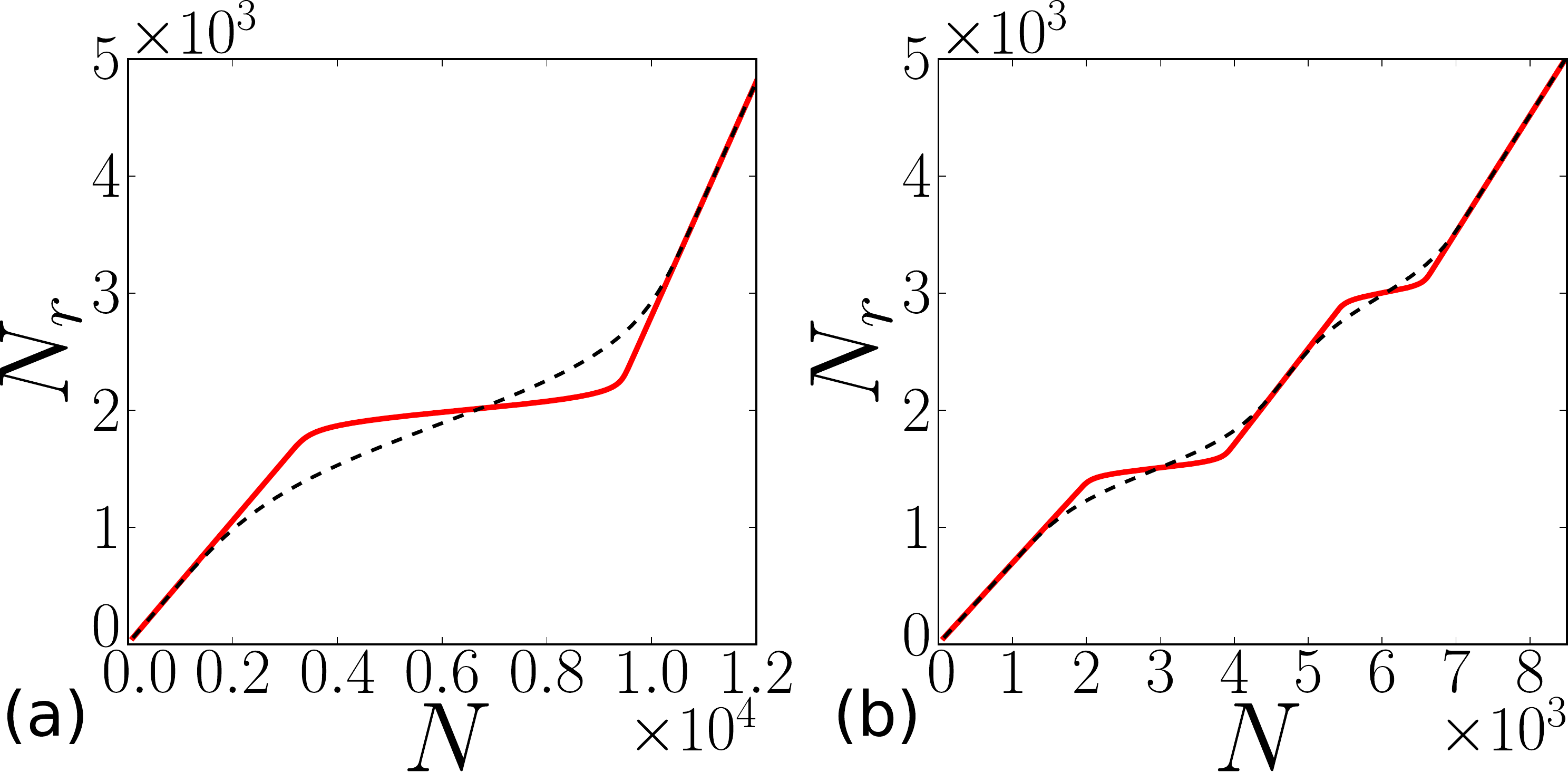}
\caption{\label{fig::7} (colour online) Particle reservoir concentration as a function of the total number of particles for normal distributed exit rates (a) and exit rates following a bimodal distribution a bimodal distribution (b). In both cases, $L=300,\, M=30, \, \alpha_0^*/V = 10^{-4}$. In panel (a) the average value of the Gaussian peak is chosen $\langle\beta\rangle=0.2$, the dashed (black) line represents the distribution with $\sigma=0.05$, the full (red) line represents the distribution with $\sigma=0.01$. In (b) the two peaks are centered at $\beta_1=0.15$ and $\beta_2=0.3$, respectively. The dashed (black) line represents the distribution for $\sigma=0.025$, the full (red) line for $\sigma=0.005$.}
\end{center}
\end{figure}
Figure~\ref{fig::7}a shows the inverse relation $N_r(N)$, for two different widths $\sigma$ of the distribution of $\beta$ values. The most striking feature is the ``quasi-plateau'' at intermediate values of $N$, which mimics the true plateaus observed for the discrete case (see for example Figure~\ref{fig::4}). As the distribution of $\beta$ values narrows (decreasing $\sigma$) this feature becomes closer to a true plateau. The fact that this ``quasi-plateau'' in $N_r(N)$ is observed for a rather generic continuous distribution of $\beta$ values suggests that the buffering of the particle reservoir by lattices entering the shock phase is a general phenomenon, with smoothing of the distribution of lattice parameters tending to ``soften'' the buffering effect. For large $N$ (which implies large $N_r$), the terms in Eq.~(\ref{r_rhoc_gauss}) which are linear in $N_r$ dominate the $\beta$-dependent terms so that $N_r(N)$ becomes linear and independent of the $\beta$-distribution.

Figure~\ref{fig::7}b shows the corresponding results for a bimodal distribution of $\beta$ values, consisting of  two Gaussian peaks. Once again, for narrow Gaussian peaks, a plateau-like form for $N_r(N)$ is recovered, but this time with two ``quasi-plateaus''. This is analogous to the case studied in Section \ref{sec::mixed}, where each lattice subpopulation produces its own range of values of $N$ over which the system is buffered.

\subsection{Relation between the distribution of boundary rates and single lattice properties}
\label{sec::slprop}

A key feature of the systems discussed in this paper is that the behavior of all the lattices is coupled via the shared particle reservoir, so that  the function  $N_r(N)$ depends on the entire distribution of lattice parameters $P(\alpha_0,\beta)$. This implies that measurements of the reservoir particle number $N_r$, or of the current on  a few lattices, as functions of $N$, contain information on the full distribution of lattice parameters $P(\alpha_0,\beta)$. In this section, we briefly sketch how such measurements could be used to compute $P(\alpha_0,\beta)$. For simplicity, we assume that the entry rate  $\alpha_0 = \alpha_0^*$ is fixed for all lattices, so that our aim is to compute the distribution of the exit rates $P(\beta)$.

We first consider the case where one is able to measure experimentally the function $N_r(N)$. In this case, $P(\beta)$ can be obtained from the derivative $dN_r/dN$: differentiating Eq.~(\ref{r_fixed_alpha}) with respect to  $N_r$ produces a linear differential equation for the cumulative probability distribution $Q(\beta) := \int_0^\beta P(\beta') d\beta'$. This differential equation is 
\begin{eqnarray}\label{eq:Q}
\frac{dN}{dN_r}&=&1+ \cfrac{\langle L \rangle M \alpha_0^*}{V}\\ \nonumber
&+&\langle L \rangle M\left[\cfrac{\alpha_0^*}{V}\left(1-2\beta\right)Q'(\beta)-\cfrac{\alpha_0^*}{V}\,Q(\beta)\right]\;,
\end{eqnarray}
where the dependence of $Q$ on $\beta$ is given by substituting $\alpha_0^* N_r/V=\beta$.
Note that here we used \emph{Leibniz' integral rule} $\frac{d}{dy} \int_0^y f(x,y) dx = f(y,y) + \int_0^y \frac{d}{dy} f(x,y) dx$, with $x=\beta', y=\alpha_0^* N_r/V$ and the integrand $f(x,y) = (1-x-y) P(x)$. Equation~(\ref{eq:Q}) depends on $dN/dN_r=1/(dN_r/dN)$ and can be solved by standard methods (e.g. the method of variation of constants). If, rather than knowing $N_r(N)$, we know the current $J_j$ on a particular lattice $j$ (in the LD phase) as a function of $N$, we can use the relation $dJ_j/dN = (dJ_j/ dN_r)(dN_r/dN)$ (since $J_j$ depends only on $N_r$ for fixed $V$, $M^{(i)}$, $\alpha_0^*$ and $\beta_j$) to write
\begin{equation*}\label{eq:conf22}
\frac{dN}{dN_r}= \left(\cfrac{dJ_j}{dN_r}\right) \big{/}\left(\cfrac{dJ_j}{dN}\right) \;. 
\end{equation*}
and note that $dJ_j/ dN_r$ can be computed from the TASEP result (for lattices in the LD phase) $J_j= \gamma \alpha_j(1-\alpha_j)$ where $\alpha_j=\alpha_0^*N_r/V$. Having thus obtained $dN/dN_r$, we can again use the derivative of Eq.~(\ref{r_fixed_alpha}) to extract $P(\beta)$. Note, however, that this procedure only works for the range of values of $\beta$ for which lattice $j$ remains in the LD phase; if the lattice is in the HD or MC phase, the current $J_j (N)$ contains no information on the reservoir particle number and cannot be used to obtain $P(\beta)$. While it remains to be seen how useful the prescription outlined here would actually be for extracting $P(\beta)$ from real (noisy) data,  this discussion highlights the important point that, in principle, one can extract information on the parameter distribution of the whole system from measurements of the behavior of just a single system component.

\section{Discussion}
\label{sec::conclusion}

In this paper, we have presented a mean-field theoretical framework to study systems in which multiple TASEPs with different parameters compete for a common pool of particles. We expect this approach to be useful in modelling a wide range of systems, from  control of gene expression in biological cells to traffic flow problems. 
Previous work has addressed the effects of a finite particle reservoir on TASEP dynamics, for single lattices ~\cite{adams_far--equilibrium_2008, cook_feedback_2009} and for multiple lattices with equal boundary rates~\cite{cook_competition_2009}); here, we extend this work to mixed populations of lattices with an arbitrarily complex distribution of parameters. 

Our theoretical approach, presented in Section~\ref{sec:theory}, is based  on combining the equation for conservation of the total number of particles with the mean-field results for the standard TASEP. This approach provides a simple way to deal with mixed populations of lattices. Although our mean-field theory does not provide information on fluctuations or on density profiles, it nevertheless reveals interesting phenomena which emerge from the competition for particles in a mixed multi-TASEP system and provides a method to calculate the full phase diagram. These phenomena arise because  the finite reservoir effectively couples all the lattices, so that a phase transition on one lattice influences the behavior of the others. Although in this work we used the mean-field TASEP results, one could easily incorporate into the same framework exact or simulated relations for the particle density as a function of the entry and exit rates.

A key observation which emerges from our work is that for a mixed population of lattices coupled to a finite reservoir of particles, any lattice subtype which  enters the SP absorbs all further particles added to the system,  buffering the  particle reservoir and making the currents and densities of all other lattices insensitive to changes in the total particle number $N$, even though these lattices may be far from their phase boundaries. This effect is specific to systems where the  lattices have {\em{different}} intrinsic entry and exit rates: if $\alpha_0$ and $\beta$ are the same for all lattices (and if lattices are large enough to neglect finite size effects), phase transitions occur on all lattices at the same critical particle reservoir number and the current does not show an additional plateau as a function of $N$. This was the case in previous work~\cite{adams_far--equilibrium_2008}, which considered a mixture of lattice with different lengths but identical boundary rates. In these conditions, a plateau in the current as a function of $N$ is due only to finite size effects.

The physical mechanism underlying the buffering of the system to changes in $N$ is attributable to  lattices undergoing the LD-HD transition that ``soak up''changes in the reservoir particle number $N_r$. As they undergo  this transition, lattices enter the shock phase, in which a queue of particles forms at the end  of the lattice. For shock phase lattices, the particle entry rate $\alpha = \alpha_0 N_r/V$ is fixed by the exit rate $\beta$  ($\alpha=\beta$): thus the number of particles on a shock phase lattice adjusts to compensate for changes in $N_r$. In the phase diagram for the standard TASEP model, the shock phase occurs only on the line separating the HD and LD phases, where $\alpha = \beta$; in the case of a finite reservoir of particles, however, the shock phase occupies a finite region of the $(\alpha_0,\beta)$ phase diagram. This is because the same entry rate $\alpha$ can be achieved over a range of $N$, $N_r$ being set by the position of the domain wall. 

An interesting analogy can be drawn between this phenomenon and first order phase separations in  (equilibrium) thermodynamics. The plateau in $N_r(N)$ which arises in our models is a direct consequence of the discontinuity in the particle density as a lattice undergoes the LD-HD phase transition. Similarly, a first order phase transition such as the boiling of water involves a discontinuity in the entropy, which is associated with latent heat: during the transition, the temperature remains constant even though further heat energy is constantly being supplied. In this analogy, heat plays the role of $N$ in our models while temperature plays the role of the reservoir particle number $N_r$.

We also show in this paper that the coupling between lattices induced by a finite particle reservoir makes it possible (under some circumstances) to  extract the entire distribution of lattice parameters from measurements of the reservoir density, or indeed the current carried by a single lattice subtype, as a function of the total particle number $N$. It will be interesting to explore the feasibility of this approach for extracting information from real, noisy, experimental data.

The multi-track TASEP with finite particle reservoir studied here bears a similarity to previous work on TASEPs on closed networks~\cite{embley_hex-tasep:_2008}. In particular, a system of $M$ lattices with a common reservoir can be mapped onto a network topology formed by $M$ rings having a unique common site ~\cite{raguin_unpublished}. The  common site plays an analogous role to a particle reservoir. However, in that problem, in contrast to the one studied here, both the  exit and entry rates depend on the occupation of the common site.

A key priority for future work must be to explore ways to include the effects of fluctuations, which are neglected in our mean-field approach. Previous work has shown that interesting fluctuation-driven effects, including localisation of domain boundaries, can occur in TASEPs with finite particle number~\cite{cook_feedback_2009}; extending this work to complex mixtures of TASEP is likely to prove fruitful. Another promising avenue  may be to study how the behavior of systems of the type studied here changes with changes in their volume: this should prove relevant when modelling transport or protein production dynamics in growing biological cells.

In summary, we have presented a simple and intuitive mean-field theoretical framework for  studying  multi-TASEP problems with finite reservoir of particles. Our method has allowed us to show that interesting physical phenomena emerge from the competition for particles among non-identical lattices, including buffering of the system to changes in the total particle number. This approach should prove a versatile tool for studying a wide variety of ``real-world'' problems~\cite{brackley_preparation} involving competition among complex populations of transport processes.

\acknowledgements
We thank Chris A.~Brackley, Michael E.~Cates, Martin R.~Evans, Marco Thiel, Ian Stansfield and the anonymous referee for valuable discussions and comments on the manuscript. RJA was supported by a Royal Society University Research Fellowship, LC by a SULSA studentship and partially by the GDRE 224 GREFI-MEFI CNRS-INdAM; PG by a DAAD postdoc fellowship and by EPSRC under grant EP/E030173, and MCR by SULSA and BBSRC (BB/F00513/X1, BB/G010722).  The collaboration leading to this work was facilitated by the StoMP research network under BBSRC grant BB/F00379X/1 and by the e-Science Institute under theme 14 ``Modelling and Microbiology''. 
\appendix
\section*{Appendix: Critical points}
In this appendix, we provide explicit expressions for the location of the critical lines for the homogeneous and mixed populations of TASEPs discussed in Sections \ref{sec::homo} and \ref{sec::mixed}.

\subsection{Homogeneous multi-track TASEP}\label{app::homo}
We first present the phase boundaries for the homogeneous case of Section \ref{sec::homo}, in which the $M$ lattices have identical boundary rates $\alpha_0$ and $\beta$. In our mean-field approach, all the lattices are in the same phase, which we denote the ``global phase'' of the system. Table \ref{tab::conditions} gives the conditions determining the global phase diagram, for the cases where the exit rate $\beta$ is limiting ($\beta < 1/2$) and where $\beta$ is not limiting ($\beta \geqslant 1/2$). As discussed in Section \ref{sec::homo}, one may  choose as variable parameters either the entry and exit rates $\alpha_0$ and $\beta$, or the number of particles $N$ and lattices $M$. 
\begingroup 
\squeezetable
\begin{table}
	\caption{\label{tab::conditions} Explicit expressions for the constraints to be satisfied in each global phase of the homogeneous multi-track TASEP with $M$ lattices of length $L$, each with intrinsic entry rate $\alpha_0$ and exit rate $\beta$, with $N$ particles. The phase boundaries are given in the case $\beta < 1/2$ (left column) and $\beta \geqslant 1/2$ (right column). Each row corresponds to a global phase (LD;HD;MC;SP) and for each phase the constraints are given in terms of  $\alpha_0$ and $\beta$ (upper line; corresponds to Figure \ref{fig::2}) and $N$ and $M$  (lower line; corresponds to Figure \ref{fig::3}).}
	\begin{ruledtabular}
		\begin{tabular}{r c  c}
			& $\beta<1/2$ & $\beta \geqslant 1/2$ \vspace{0.5ex}\\
			\hline \\[-1.5ex]
			\multirow{5}{*}{\textbf{LD}} & $\cfrac{\alpha_o}{V} < \cfrac{\beta}{N - L M \beta}$ &  $\cfrac{\alpha_o}{V} < \cfrac{1}{2N - L M}$\\ [1.0ex] \\
			 & $N < V \cfrac{\beta}{\alpha_o} + L M \beta$ & $N < \cfrac{V}{2 \alpha_o} + \cfrac{L M}{2}$\\[4.5ex]

			\hline \\[-1.5ex]
			\multirow{5}{*}{\textbf{HD}}& $\cfrac{\alpha_o}{V} > \cfrac{\beta}{N - L M (1-\beta)}$ &  $-$ \\[2.5ex]
								 & $N > V \cfrac{\beta}{\alpha_o} + L M (1-\beta)$ & $-$\\[4.5ex]

			\hline \\[-1.5ex]
			\multirow{3}{*}{\textbf{MC}}	& $-$ &  $\cfrac{\alpha_o}{V} \geqslant \cfrac{1}{2N - L M}$ \\ [2.5ex]
																		& $-$ & $N  \geqslant \cfrac{V}{2 \alpha_o} + \cfrac{L M}{2}$\\[2.5ex]

			\hline \\[-1.5ex]
			\multirow{5}{*}{\textbf{SP}} & $ \cfrac{\beta}{N - L M\beta} < \cfrac{\alpha_o}{V} < \cfrac{\beta}{N - L M (1-\beta)}$ &  $-$ \\[3.5ex]
			& $ V \cfrac{\beta}{\alpha_o} + L M \beta < N < V \cfrac{\beta}{\alpha_o} + L M (1-\beta)$ & $-$\\[2.5ex]

		\end{tabular}
	\end{ruledtabular}
\end{table}
\endgroup 
These alternative choices correspond respectively to the phase diagrams shown in Figures \ref{fig::2} and \ref{fig::3}. Table \ref{tab::conditions} gives explicit forms for the phase boundaries in both these cases: the upper line in each row gives the conditions on $\alpha_0$ and $\beta$ (assuming fixed $N$ and $M$), while the lower line gives the conditions on $N$ and $M$ (for fixed $\alpha_0$ and $\beta$).

\subsection{Mixed population of TASEPs}
\label{app::cirit_points}
For the mixed multi-track TASEP discussed in Section \ref{sec::mixed}, which is composed of two subpopulations of lattices with different intrinsic entry rates $\alpha_0$, the states of the system are characterised by the phases of the two lattice subpopulations (e.g. in the LD/SP state, subpopulation 1 is in the low density phase while subpopulation 2 is in the shock phase). Boundaries between these states  occur when the $\theta$-functions in Eq.~(\ref{eq::r_two_types_theta}) are approached from above or below (at these points one of the lattice subpopulations undergoes a phase transition). More precisely, if $\beta<1/2$ the phase transitions are located at:
\begin{eqnarray}
	\label{eq::crit_points1}
	\text{LD/LD} \rightarrow \text{LD/SP}:  & \displaystyle \lim_{N_r  \to \left(V\beta/\alpha_{0}^{(2)}\right)^-} & N(N_r) \\ \nonumber
	\text{LD/SP} \rightarrow \text{LD/HD} : &  \displaystyle \lim_{N_r  \to \left(V\beta/\alpha_{0}^{(2)}\right)^+} & N(N_r) \\ \nonumber
	\text{LD/HD} \rightarrow \text{SP/HD} : &  \displaystyle \lim_{N_r  \to \left(V\beta/\alpha_{0}^{(1)}\right)^-} & N(N_r) \\ \nonumber
	\text{SP/HD} \rightarrow \text{HD/HD} : &  \displaystyle \lim_{N_r  \to \left(V\beta/\alpha_{0}^{(1)}\right)^+} & N(N_r) \;,
\end{eqnarray}
while if $\beta \geqslant 1/2$:
\begin{eqnarray}
	\label{eq::crit_points2}
	\text{LD/LD} \rightarrow &\text{LD/MC}:  \displaystyle \lim_{N_r  \to V/2\alpha_{0}^{(2)}} & N(N_r) \\ \nonumber
	\text{LD/MC} \rightarrow &\text{MC/MC} : \displaystyle \lim_{N_r  \to V/2\alpha_{0}^{(2)}} & N(N_r) \;.
\end{eqnarray}
Tables~\ref{table::crit_points1} and~\ref{table::crit_points2} give explicit expressions for the constraints on the parameters $N,L,V,M^{(1)},M^{(2)},\alpha_{0}^{(1)}, \alpha_{0}^{(2)}$ and $\beta$, for each of the possible states of the system, in the cases where $\beta < 1/2$ (Table~\ref{table::crit_points1}) and $\beta \geqslant 1/2$ (Table~\ref{table::crit_points2}). To obtain these phase boundaries we insert the limits defined in Eqs.~(\ref{eq::crit_points1}) and (\ref{eq::crit_points2}) into Eq.~(\ref{eq::r_two_types_theta}), using $\lim_{z  \to 0^+} \theta(z)= 1$ and $\lim_{z  \to 0^-} \theta(z)= 0$. Note that this simply means to substitute the critical values $N_r = \beta V/\alpha_0^{(i)}$ (LD-HD transition) and $N_r = V/2\alpha_0^{(i)}$ (MC transitions) respectively, while the limit from below corresponds to substituting the $\theta$-function in Eq.~(\ref{eq::r_two_types_theta}) with $\theta(z)=0$ for the lower limit and $\theta(z)=1$ for the upper limit.
\begingroup 
\squeezetable
\begin{table} []
	\caption{Phase boundaries for the multi-track TASEP with two lattice subpopulations introduced in Sec.~\ref{sec::mixed}, for the case $\beta < 1/2$, obtained by combining Eqs.~(\ref{eq::crit_points1}) with (\ref{eq::r_two_types_theta}). \label{table::crit_points1}}
	\begin{ruledtabular}
		\begin{tabular}{l c}
			\\[-1.5ex]
			\textbf{\large{Low Density/Low Density}} \\[2.5ex]
			$N <  \cfrac{V\beta}{\alpha_{0}^{(2)}} + \beta \cfrac{\alpha_{0}^{(1)}}{\alpha_{0}^{(2)}} L M^{(1)} + \beta L M^{(2)}$ \\[2.5ex]
			\hline \\[-1.5ex]
			
			\textbf{\large{Low Density/Shock Phase}} \\[2.5ex]
					$\cfrac{V \beta}{\alpha_{0}^{(2)}} + \beta \cfrac{\alpha_{0}^{(1)}}{\alpha_{0}^{(2)}} L M^{(1)} + \beta L M^{(2)} < N <\cfrac{V \beta}{\alpha_{0}^{(2)}} + \beta \cfrac{\alpha_{0}^{(1)}}{\alpha_{0}^{(2)}} L M^{(1)} + (1-\beta)L M^{(2)}$\\[2.5ex]
			\hline \\[-1.5ex]
			
			\textbf{\large{Low Density/High Density}} \\[2.5ex]			$\cfrac{V \beta}{\alpha_{0}^{(2)}} + \beta \cfrac{\alpha_{0}^{(1)}}{\alpha_{0}^{(2)}} L M^{(1)} + (1-\beta)L M^{(2)} < N <\cfrac{V \beta}{\alpha_{0}^{(1)}} + \beta L M^{(1)} + (1-\beta)L M^{(2)}$\\[2.5ex]
			\hline \\[-1.5ex]
			
			\textbf{\large{Shock Phase/High Density}} \\[2.5ex]
			$ \cfrac{V\beta}{\alpha_{0}^{(1)}} + \beta L M^{(1)} + (1-\beta)L M^{(2)} < N <V \cfrac{\beta}{\alpha_{0}^{(1)}} + (1-\beta)(M^{(1)}+M^{(2)}) L $\\[2.5ex]
			\hline \\[-1.5ex]

			\textbf{\large{High Density/High Density}} \\[2.5ex]
			$N >\cfrac{V\beta}{\alpha_{0}^{(1)}} + (1-\beta)(M^{(1)}+M^{(2)}) L $\\[2.5ex]

			\end{tabular}
	\end{ruledtabular}
\end{table}
\endgroup 

\begingroup 
\squeezetable
\begin{table} [htbp]
	\caption{Phase boundaries for the multi-track TASEP with two lattice subpopulations introduced in Sec.~\ref{sec::mixed}, for the case $\beta \geqslant 1/2$, obtained by combining Eqs.(\ref{eq::crit_points2}) with (\ref{eq::r_two_types_theta}). \label{table::crit_points2}}
	\begin{ruledtabular}
		\begin{tabular}{l c}
			\\[-1.5ex]
			\textbf{\large{Low Density/Low Density}} \\[2.5ex]
			$N <  \cfrac{V}{2\alpha_{0}^{(2)}} + \cfrac{\alpha_{0}^{(1)}}{2\alpha_{0}^{(2)}} L M^{(1)} + \cfrac{L M^{(2)}}{2}$ \\[2.5ex]
			\hline \\[-1.5ex]
			
			\textbf{\large{Low Density/Maximal Current}}& \\[2.5ex]
			$\cfrac{V}{2\alpha_{0}^{(2)}} + \cfrac{\alpha_{0}^{(1)}}{2\alpha_{0}^{(2)}} L M^{(1)} + \cfrac{L M^{(2)}}{2} < N <  \cfrac{V}{2\alpha_{0}^{(1)}} + \cfrac{L (M^{(1)}+M^{(2)})}{2}$\\[2.5ex]
			\hline \\[-1.5ex]
			
			\textbf{\large{Maximal Current/Maximal Current}}& \\[2.5ex]
			$N >  \cfrac{V}{2\alpha_{0}^{(1)}} + \cfrac{L (M^{(1)}+M^{(2)})}{2}$ \\[2.5ex]
			\end{tabular}
	\end{ruledtabular}
\end{table}
\endgroup


\end{document}